# OPTIMIZING INTRUSION DETECTION SYSTEM PERFORMANCE THROUGH SYNERGISTIC HYPERPARAMETER TUNING AND ADVANCED DATA PROCESSING

Samia Saidane [1], Francesco Telch [2], Kussai Shahin [2] and Fabrizio Granelli [1]

[1]DISI, University of Trento, Trento, Italy
[2]Trentino Digitale Spa, Trento, Italy

## ABSTRACT

*Intrusion detection is vital for securing computer networks against malicious activities. Traditional methods struggle to detect complex patterns and anomalies in network traffic effectively. To address this issue, we propose a system combining deep learning, data balancing (K-means + SMOTE), high-dimensional reduction (PCA and FCBF), and hyperparameter optimization (Extra Trees and BO-TPE) to enhance intrusion detection performance. By training on extensive datasets like CIC IDS 2018 and CIC IDS 2017, our models demonstrate robust performance and generalization. Notably, the ensemble model "VGG19" consistently achieves remarkable accuracy (99.26% on CIC-IDS2017 and 99.22% on CSE-CIC-IDS2018), outperforming other models.*

## 1. INTRODUCTION

With the rapid development of computer and communications networks, Internet technology has provided more convenient services to people around the world than ever before. However, the escalating number and diverse range of cyberattacks, including network viruses, malicious eavesdropping, and targeted malicious attacks, pose significant risks to individuals' information security and the safety of their assets. These growing threats demand utmost attention to safeguard people's personal information and property. These risks can be mitigated to an extent by employing a network intrusion-detection system (NIDS) that can identify malicious activities and potential security breaches in real time [1].

The task of network intrusion detection aims at discovering suspicious attacks while taking the corresponding measures to protect the network from attacks and reduce consequent economic losses. Network intrusion detection has become an important research content in the field of network security [1] [2]. The traditional Network Intrusion Detection Systems (NIDSs) commonly rely on two main detection methods: misuse detection and anomaly detection [3]. Anomaly detection is centred on recognizing deviations from typical network behaviour by analyzing patterns and employing statistical models. Misuse detection involves identifying known attack patterns or signatures, making it effective in detecting specific predefined threats but less adaptable to novel or evolving attack strategies. However, these methods have inherent drawbacks, including a low detection rate and a high false-positive rate, as indicated by previous studies [3]. Although widely utilized, these approaches face limitations in accurately detecting sophisticated and evolving cyber threats [4] [5].

As cyberattacks continue to grow in complexity, researchers and practitioners have introduced a range of machine learning (ML) and deep learning (DL) techniques to enhance the effectiveness of Network Intrusion Detection Systems (NIDSs). The integration of artificial intelligence (AI) in Intrusion Detection Systems (IDS) has emerged as a significant area of interest, as highlighted in [5]. However, despite advances in the use of artificial intelligence (AI) for intrusion classification in Intrusion Detection Systems (IDS), there are several major challenges that need to be addressed in the design and implementation of IDS:

- When dealing with large-scale, high-dimensional data points, traditional network intrusion detection approaches tend to apply dimension reduction to remove noise in measurements. Consequently, they are likely to remove significant information when extracting features for intrusion behaviours resulting in high false detection rate.

- The challenge of imbalanced data is a common issue when constructing a network intrusion detection model using deep learning. Such data can impact the model's performance, resulting in elevated false alarm rates and increased false miss rates for certain minority class samples.

- The characteristics of network traffic data are very complicated, which make it difficult to extract features. If the right data features cannot be fully extracted, the classification results will be poor.

- Deep learning models often involve multiple hyperparameters that require tuning to attain optimal performance. The process of exhaustively searching through all potential combinations of these hyperparameters can be both computationally expensive and time-consuming.

The rest of the paper is organized as follows. Section 2 presents a review of previous relevant works. The proposed methodology for intrusion detection systems is detailed in Section 3. Experimental evaluation results are presented in Section 4, and Section 5 provides conclusions and future works.

## 2. RELATED WORKS

The increasing volume of network-generated data and advancements in machine learning have prompted the development of automatic and adaptive solutions to tackle intrusion detection problems. Various machine learning and deep learning approaches have been proposed to address intrusion detection issues [5]. Researchers have leveraged machine learning to detect a wide range of attacks [6]. Ünal Çavuşoğlu et al. suggested a layered architecture approach that selects appropriate machine learning algorithms based on the attack type [7]. Commonly studied algorithms include Support Vector Machine (SVM), Naive Bayes, Random Forest (RF), and other clustering techniques. Zhao introduced the least squares support vector machine (LSSVM) algorithm with a hybrid kernel function and optimized each LSSVM parameter using particle swarm optimization (PSO) [8]. Thaseen et al. proposed a model using Chi-square feature selection and integrated SVM, Modified Naïve Bayes (MNB), and LPBoost to make predictions based on majority voting [9]. Sumaiya et al. devised an intrusion detection model employing Chi-square feature selection and a multi-class support vector machine, resulting in enhanced detection rates and decreased occurrences of false alarms [10].

Tao et al. introduced the FWP-SVM-GA model, which involves feature selection, weight optimization, and parameter tuning for support vector machines using genetic algorithms. This approach demonstrated improved detection rates and reduced false positives and false negatives compared to other SVM-based intrusion detection methods [11]. Peng et al. presented a clustering methodology for intrusion detection systems by incorporating Mini Batch K-means and Principal Component Analysis (PCA). This approach addresses the challenge of high-dimensionality in datasets, offering a solution to effectively handle such complex data structures. This method is suitable for intrusion detection in big data environments [12].

In the same context, Deep Learning (DL) has demonstrated effectiveness across diverse domains, including its notable applications in the realm of cybersecurity and intrusion detection systems [13]. Yin et al. introduced a deep learning method for intrusion detection utilizing Recurrent Neural Networks (RNN). Their study explored the model's performance in binary and multi class classification, investigating the influence of neuron count and learning rate on the model's efficacy [14]. Kim applied the Long Short-Term Memory (LSTM) architecture within an RNN framework and verified the effectiveness of deep learning for Intrusion Detection Systems (IDS) through performance testing [15].

These investigations encourage the utilization of deep neural networks to comprehend the hierarchical features of network traffic, encompassing both spatial and temporal aspects, to classify network traffic effectively [16]. However, the substantial imbalance in network intrusion traffic results in significant variations in the proportions of various traffic data. Most detection methods primarily focus on diminishing the overall average false positive rate, which, unfortunately, tends to elevate the false positive rate specifically in minority samples. In a real network environment, the attacks of minority attacks are more dangerous than those of majority attacks. To tackle this issue, Bamakan et al. introduced a precise and robust approach based on Ramp Loss K-Support Vector Classification-Regression (K-SVCR) to handle highly imbalanced and skewed attack distributions in multi-class intrusion detection [17]. Yan et al. proposed a traffic classification method to balance imbalanced network data by introducing an enhanced Synthetic Minority Over-sampling Technique (SMOTE) known as Mean SMOTE (M-SMOTE) [18]. However, the performance of those deep learning-based Network Intrusion Detection System (NIDS) relies significantly on the selection of appropriate hyperparameters. These hyperparameters include the number of layers, filter size, activation functions, and learning rate [19]. The selection of optimal hyperparameters is crucial as it directly affects the accuracy and generalization ability of the deep learning model used for intrusion detection. To address these issues, a randomized search approach for hyperparameter optimization in IDS models was proposed in [20]. This technique involved randomly sampling hyperparameters from a predefined search space and evaluating their performance on a validation set, efficiently utilizing computational resources to determine improved model parameters for deep convolutional neural networks (DCNNs).

In reference [21], a distinctive approach was introduced, employing lion swarm optimization to fine-tune the hyperparameters of the proposed optimized CNN hierarchical multiscale long short-term memory (OCNN-HMLSTM) model. Nonetheless, resolving a specific issue in intrusion detection systems might inadvertently give rise to other challenges, resulting in a gap in attaining a comprehensive and universally applicable solution.

In this paper, we introduce a novel intrusion detection solution designed to overcome challenges associated with NetFlow data imbalance, feature engineering, high dimensionality, and hyperparameter tuning. Our proposed intrusion detection system distinguishes itself through a holistic integration of various techniques, aiming to significantly enhance the performance and effectiveness of intrusion detection systems, thereby improving network security in real-world scenarios. The key contributions of our proposed intrusion detection system are as follows:

- Addressing NetFlow Data Imbalance: The system effectively tackles the challenge of NetFlow data imbalance by implementing data balancing techniques, specifically leveraging the Hybrid Kmean+SMOTE method. This approach significantly enhances the balance in class distribution within the dataset, thereby improving the overall performance of intrusion detection models.

- Handling High Dimensionality with PCA: To manage the high dimensionality inherent in NetFlow data, the system utilizes Principal Component Analysis (PCA). By reducing the dimensionality while preserving crucial information, PCA enables more efficient processing and analysis of the data, contributing to the system's effectiveness.

- Feature Selection Using FCBF Technique: The system incorporates the Fast Correlation-Based Filter (FCBF) feature selection technique. FCBF plays a pivotal role in identifying and selecting the most informative features relevant to intrusion detection.

- Optimizing Hyperparameters with BO-TPE: The proposed system employs Bayesian optimization, specifically utilizing a tree-based Parzen estimator (BO-TPE), to fine-tune the hyperparameters of deep learning models. This optimization process not only improves the accuracy of the models but also enhances their generalization ability.

## 3. METHODOLOGY

The proposed network intrusion detection system, as illustrated in Figure 1, follows a systematic flow consisting of several stages. Initially, an initial phase of data analysis and cleaning is conducted to ensure the quality and integrity of the dataset. Following this, the data undergoes numerical normalization and other pre-processing steps to prepare it for further analysis. A feature engineering technique, Feature selection by Fast Correlation Based Filter (FCBF), is then employed to select the most relevant features for intrusion detection. Subsequently, the data is subjected to high-dimensional reduction using Principal Component Analysis (PCA) to reduce dimensionality and capture the most significant information. To address the issue of imbalanced data, a Hybrid data sampling processing technique is applied, which helps achieve a balanced dataset.

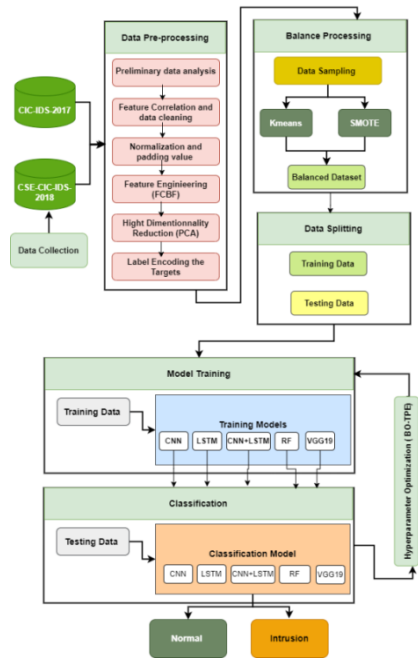

Figure 1. Proposed Simulation Methodology

Finally, the balanced data is split into training and testing sets and used to train five separate deep learning models. These models are specifically designed for network traffic classification, distinguishing between normal and anomalous instances. The BO-TPE optimizer is used to fine-tune the deep learning models systematically. Leveraging the power of deep learning and the insights gained from the previous stages, the proposed network intrusion detection model aims to enhance the accuracy and effectiveness of intrusion detection in real-world scenarios.

### 3.1. Datasets

#### 3.1.1. CIC-IDS2017 Dataset

The CIC-IDS 2017 dataset, short for Canadian Institute for Cybersecurity Intrusion Detection System 2017 dataset, is a comprehensive and widely used cybersecurity dataset. It contains network traffic data collected in a controlled lab environment, simulating real-world network traffic and attacks. This dataset encompasses various network traffic features, including packet-level data, flow-level statistics, and higher-level protocol information. CIC-IDS2017 introduces eleven new attacks, including Brute Force, PortScan, DoS, web attacks like XSS and SQL Injection, FTP-Patator, and SSH-Patator. Developed in 2017 by the Canadian Institute for Cybersecurity, this dataset, with its eighty features, is used to monitor both benign and malicious network traffic.

#### 3.1.2. CSE-CIC-IDS 2018 Dataset

The CSE-CIC-IDS2018 dataset, developed by the Canadian Institute for Cybersecurity (CIC), is a significant and reliable dataset specifically created for assessing and benchmarking Intrusion Detection Systems (IDS) within the cybersecurity domain. It is widely recognized as a valuable resource for evaluating intrusion detection models that focus on network anomalies [21]. This dataset encompasses a diverse range of attacks, including ten distinct classes, and is organized based on the relative occurrence of these attacks in the data. The classes include Benign, Bot,

FTP-BruteForce, SSH-Bruteforce, DDOS attack-HOIC, DDOS attack-LOIC-UDP, DoS attacks-GoldenEye, DoS attacks-Slow HTTP test, intrusion, and web attacks. The CSE-CIC-IDS2018 dataset plays a pivotal role in facilitating research, experimentation, and the assessment of Intrusion Detection Systems [12].

### 3.2. Data Pre-Processing

In this section, our goal is to prepare the data in a manner that makes it immediately suitable for training the model. The following stages were implemented to process the data before commencing the module training.

#### 3.2.1. Preliminary Data Analysis

To gain insights into this dataset, we conduct a preliminary data analysis, focusing on the distribution of attacks and the generation of statistical plots. Initially, we examine the distribution of attacks within the dataset, which includes 15 types of attacks, such as DDoS, DoS, Bot, Brute Force, and Web Attacks. This analysis helps us understand the prevalence and frequency of different attack categories within the dataset. Furthermore, we generate statistical plots, such as histograms, box plots, and bar plots, to visualize the characteristics of the dataset. These plots assist in showcasing the distribution of features, identifying outliers, and comprehending the overall structure of the data.

#### 3.2.2. Feature Correlation and Data Cleaning

In intrusion detection with NetFlow data, the initial phase includes feature correlation and data cleaning for accurate analysis and modelling. This process involves tasks column renaming, null value removal, non-finite value removal, and the creation of univariate histograms. The workflow starts with loading the NetFlow dataset and then thoroughly examining feature correlations using correlation coefficients or statistical methods. Univariate histograms are generated to visualize feature distributions. These steps are essential for preparing the NetFlow CIC-IDS datasets for precise and effective intrusion detection analysis.

#### 3.2.3. Normalization and Padding Value

A substantial disparity in feature data dimensions within the dataset can lead to issues such as sluggish model training and negligible improvements in accuracy. To address this concern, the dataset was subjected to mapping using the MinMaxScaler [24], bringing the data within the range of (0,1):

$$X_{\text{normalized}} = \frac{X - X_{min}}{X_{max} - X_{min}} \qquad (1)$$

where $X_{max}$ is the maximum value, and $X_{min}$ is the minimum value.

#### 3.2.4. Feature Engineering

Using Fast Correlation-Based Filter (FCBF) for feature engineering in the NetFlow CIC-IDS dataset involves systematically selecting the most relevant features for intrusion detection. This process starts by computing individual feature-class and feature-feature correlation scores. The top-ranked features with strong correlations to class labels are retained, while redundant features are removed, ensuring a diverse set of informative features. FCBF optimizes the feature subset, maximizing relevance to the target class while minimizing redundancy. Algorithm 1 provides a detailed overview of the FCBF procedure within IDS datasets, enhancing intrusion detection efficiency and accuracy.

---

**Algorithm1** Fast Correlation-Based Filter (FCBF) for Feature Selection on CIC-IDS Net Flow Dataset

**Require:** CIC-IDS Net Flow Dataset $D$, Threshold $T$
**Ensure:** Selected feature set $F$
1: Compute correlation between each feature and the target variable (intrusion vs. non-intrusion)
2: Sort features based on correlation in descending order
3: Initialize empty feature set $F$
4: **for** each feature $f$ in sorted features **do**
5:   Compute merit of feature $f$ using FCBF formula adapted for CIC-IDS Net Flow dataset
6:   **if** merit of $f >$ Thre s hold $T$ **then**
7:     Add feature $f$ to $F$
8:     Remove features highly correlated with $f$ from sorted features
9:   **end if**
10: **end for**
11: **return** $F$

---

### 3.2.5. Height Dimensionality Reduction

Performing high-dimensionality reduction on the NetFlow datasets using Principal Component Analysis (PCA) involves a systematic process to reduce the dataset's feature space while retaining essential information. In this study, the NetFlow dataset, with its high-dimensional feature set representing network traffic characteristics, is first collected, and pre-processed to handle missing values and normalize the data. Subsequently, PCA is applied, by computing the covariance matrix of the features and obtaining its eigenvectors and eigenvalues. These eigenvectors represent the principal components, and by selecting the top eigenvectors based on their corresponding eigenvalues, a new lower-dimensional feature space is constructed. The original features are then projected onto this reduced feature space, effectively achieving dimensionality reduction while preserving the most significant information. Figures 2 and 3 visualize the results of high-dimensionality reduction of the CIC-IDS2017 and CSE-CIC-IDS2018 datasets using PCA (Principal Component Analysis).

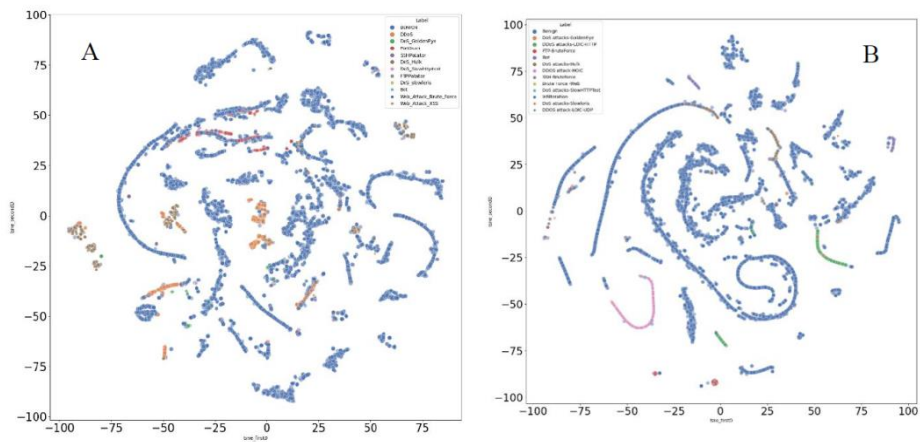

Figure 2. A: PCA CIC-IDS2017 dataset Distribution and B: PCA CSE-CIC-IDS2018 dataset Distribution

### 3.2.6. Encoding the Labeled Data

In the context of the NetFlow CIC-IDS dataset with 15 distinct classes, encoding the labelled data is crucial for making these categorical labels suitable for machine learning and deep learning algorithms. This process involves assigning unique numerical identifiers to each class, essentially converting labels into discrete numerical values. For multi-class scenarios, one-hot encoding is often utilized. This method involves encoding categorical labels with numeric values (e.g., 1, 2, 3, 4). One-hot encoding retains the categorical nature of the labels and is vital for training supervised learning models, including neural networks, decision trees, and various machine learning algorithms.

## 3.3. Kmean + Smote Data Balancing

In this paper we utilized K-means clustering in conjunction with SMOTE (Synthetic Minority Over-sampling Technique) for balancing NetFlow data. Initially, we apply K-means clustering to partition the NetFlow data into clusters. Subsequently, we identified the minority class instances, typically representing intrusions, within each cluster. Then we employed SMOTE to over sample these minority instances, generating synthetic samples to balance the dataset. The SMOTE algorithm involves selecting a minority instance and its k-nearest neighbors, then creating convex combinations of the instance and its neighbors to form synthetic data points. This process is repeated until the desired balance is achieved. The combined approach of K-means clustering, and SMOTE oversampling contributes to a more robust and balanced dataset. Algorithm 2 outlines the process of integrating K-means clustering with SMOTE for data balancing within our intrusion detection system:

---
**Algorithm 2** K Means Combined with SMOTE for Data Imbalance Processing in Intrusion Detection

---
**Require:** Net Flow data $X$, Number of clusters $k$, SMOTE factor $N$
**Ensure:** Balanced data set $X_{balanced}$
    K Means SMOTE $X$, $k$, $N$
1: Apply K-means clustering to $X$ with $k$ clusters: $X_{clusters} \leftarrow$ K Means Clustering $(X,k)$
2: Initialize empty set $X_{balanced}$
3: **for** each cluster $C$ in $X_{clusters}$ **do**
4:    Identify minority in stances within $C$(e.g., network intrusions)
5:    Apply SMOTE to oversample minority instances in $C$ with factor $N$: $X_{SMOTE} \leftarrow$ SMOTE (minority instances in $C,N$)
6:    Add $X_{SMOTE}$ to $X_{balanced}$
7: **end for**
8: **Return** $X_{balanced}$

---

## 3.4. Data Splitting

Our IDS systems dataset's have been divided into an 80% training set and a 20% testing set. Additionally, within the training set, we further divide it into training and validation subsets. This division allows us to tune our hyperparameters during training, thus enhancing the model's performance.

## 3.5. Deep Learning Models

We implemented and evaluated four deep learning models and one machine learning algorithm, which have demonstrated good performance and high accuracy in the state-of-the-art intrusion detection literature.

### 3.5.1. CNN Model

In this paper, we proposed a multi-layer Convolutional Neural Network (CNN) tailored for intrusion detection. The input layer is designed to receive a 1D sequence of NetFlow data features, characterized by an input shape of (sequence length, 1). Subsequent Conv1D layers employ learnable filters to extract spatial features from the input sequence, with the number of filters and kernel sizes as optimization hyperparameters. MaxPooling1D layers follow, down sampling the sequence while preserving essential information, and a Flatten layer reshapes feature maps into a 1D vector for seamless connectivity with dense layers. The dense layers, or fully connected layers, perform higher-level reasoning on features, incorporating ReLU activation functions for nonlinearity. Dropout layers are strategically placed after dense layers to mitigate overfitting, and the output layer employs the Softmax activation function for multiclass classification, addressing 15 attack types. Hyperparameter tuning, crucial for optimal model performance, involves adjusting the learning rate, batch size, and number of epochs. Learning rate influences parameter changes during training, impacting convergence speed and efficiency, while the batch size balances computational efficiency and model quality. The number of epochs prevents underfitting or overfitting. To effectively tune these hyperparameters, we employ Bayesian optimization with a tree-based Parzen estimator (BO-TPE).

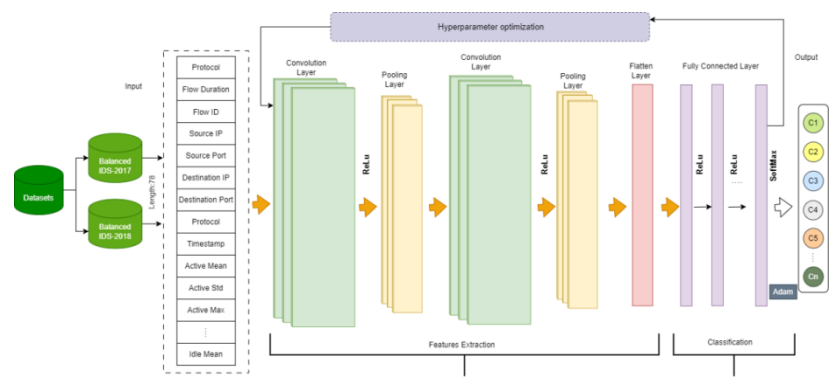

Figure 3. CNN architecture.

### 3.5.2. LSTM Model

Our approach utilizes an LSTM network architecture for multi-step prediction, consisting of LSTM layers, dropout layers, and a fully connected layer, forming a seven-layer network. This configuration enables the LSTM network to generate predictions for future time series values. In the final segment, we incorporate the local average algorithm with adaptive parameters to generate anomaly detection results based on the earlier prediction outcomes. Additionally, our approach incorporates BO-TPE for hyperparameter optimization, ensuring the effectiveness of the model in intrusion detection.

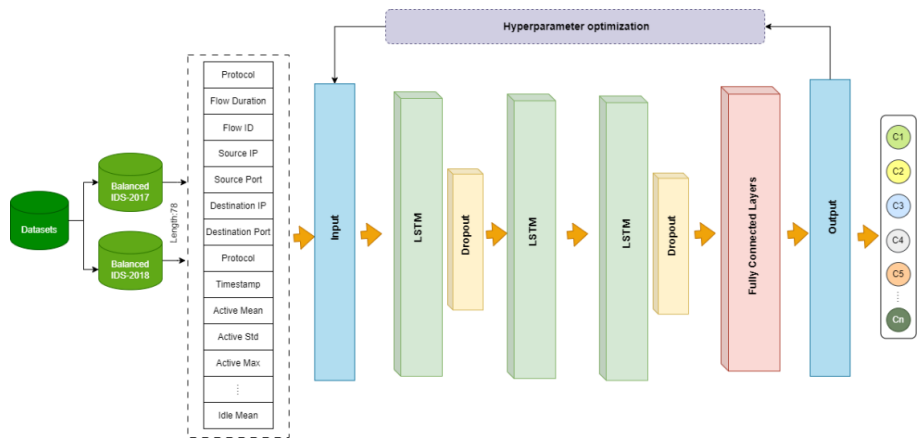

Figure 4. LSTM Architecture

### 3.5.3. CNN - LSTM Model

Our approach adeptly combines the spatial feature extraction capability of CNN with LSTM's proficiency in capturing temporal characteristics. The model begins with the CNN layer, where convolution operations extract crucial features to generate a feature map. Subsequently, the output undergoes max pooling to retain dominant spatial features, and batch normalization ensures seamless processing. The data then flows into an LSTM layer to extract temporal characteristics, with a dropout layer mitigating overfitting. This CNN-LSTM combination is iterated three times, varying neurons, and filters, and concludes with a fully connected layer using SoftMax activation for classification. By employing three kernels in the convolutional layer and the tanh activation function for data transformation, complex features are captured. Two-kernel max pooling reduces feature dimensions, and these features are mapped to the LSTM layer for extracting temporal information. Once LSTM acquires temporal insights, fusion features are directed to a fully connected layer for classification. The SoftMax activation effectively identifies and classifies attacks within NetFlow network data. This model is finely tuned for intrusion detection, with a focus on the balanced NetFlow dataset, and its hyperparameters are optimized using Bayesian optimization with a tree-based Parzen estimator (BO-TPE). Figure 5 depicts the structure of this combination.

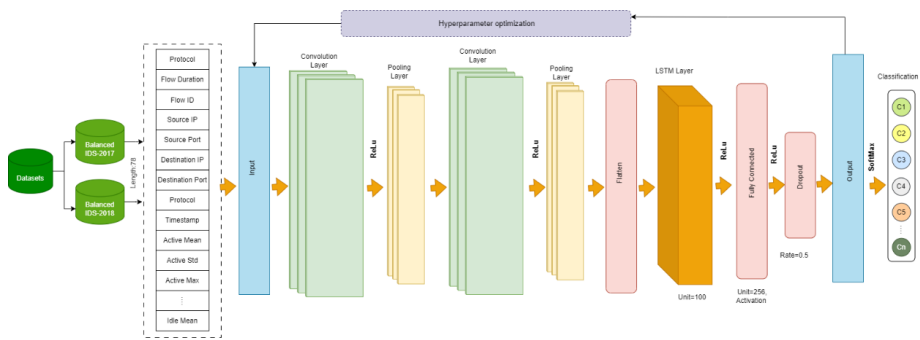

Figure 5. CNN+LSTM Architecture

### 3.5.4. VGG19 Model

The VGG19 model is a deep convolutional neural network (CNN) architecture that has been widely used for image classification tasks. In the context of intrusion detection using the Net Flow balanced dataset, The model's convolutional layers analyze spatial patterns or correlations in the data, while max-pooling layers down sample and retain essential spatial features. After flattening the data, fully connected layers capture complex relationships among the features, aiding in intrusion classification. Dropout layers mitigate overfitting, and the output layer, typically using Softmax activation, classifies different intrusion types. This adaptation requires a meticulous preprocessing of network traffic data and hyperparameter optimization, as VGG19's original design was for images. Ultimately, VGG19 can serve as a feature extractor, capturing high-level network traffic representations and improving intrusion detection by processing the data through its layers. To enhance the performance of the VGG19 model, Bayesian optimization with tree-based Parzen estimator (BO-TPE) is employed as a hyperparameter tuning technique.

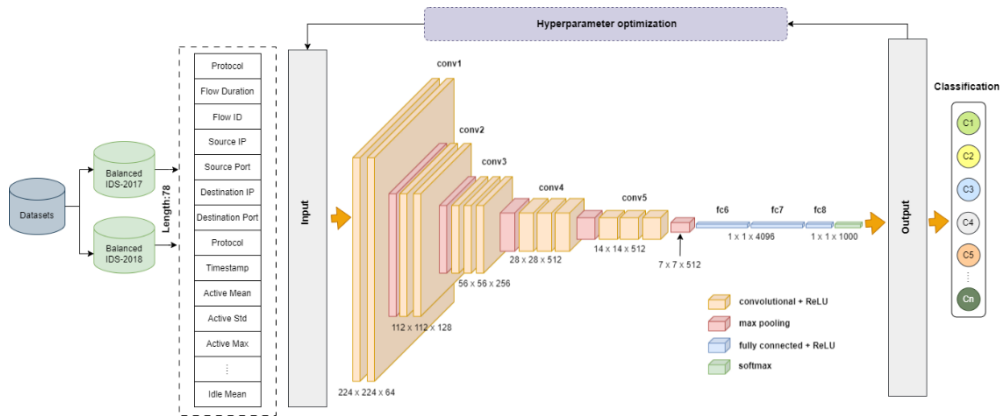

Figure 6. VGG19 Architecture

### 3.5.5. Random Forest

The Random Forest model, a widely used ensemble learning algorithm, is an effective choice for intrusion detection with the CIC-IDS NetFlow balanced dataset. This model is trained on the dataset containing NetFlow records, aiming to accurately classify network traffic into one of the 15 intrusion classes or benign. To boost its performance, Bayesian optimization with a tree-based Parzen estimator (BO-TPE) is applied to tune the model's hyperparameters. This approach combines machine learning and automation to empower the Random Forest model in identifying and categorizing various network intrusions, strengthening overall network security and integrity.

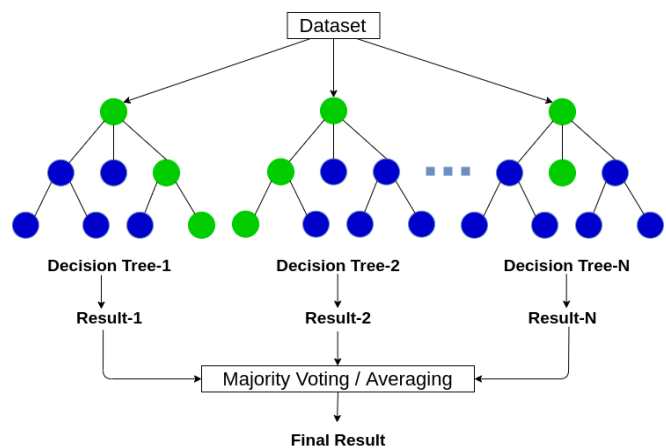

Figure 7. Random Forest Architecture

## 4. RESULTS AND DISCUSSION

In this section, we conduct a comprehensive evaluation of various deep learning algorithms within the context of our intrusion detection system. Additionally, we investigate the impact and effectiveness of Kmeans-SMOTE, Principal Component Analysis (PCA), and hyperparameter tuning across a range of deep learning algorithms.

### 4.1. Evaluation Metrics

In this paper, we utilize accuracy, precision, recall, and F1-Measure as the key performance indicators for evaluating the model. These indicators are derived from the four basic attributes of the confusion matrix: Ture Positive (TP) (attack data that is correctly classified as an attack); False Positive (FP) (normaldata that is incorrectly classified as an attack); True Negative(TN) - (normal data that is correctly classified as normal); False Negative (FN) - (attack data that is incorrectly classified as normal). To evaluate the performance of our solution wewill use the following formulas:

$$Accuracy\ (AC) = \frac{TP+TN}{TP+TN+FP+FN} \quad (2)$$

$$Precision = \frac{TP}{TP+FP} \quad (3)$$

$$Recall = \frac{TP}{TP+FN} \quad (4)$$

$$F1 - Measure = \frac{2\times Precision \times Recall}{Precision+Recall} \quad (5)$$

### 4.2. Results on CIC-IDS2017 Dataset

Table 1 and 2 provide a detailed analysis of the performance results for intrusion detection classification using the CIC-IDS2017 dataset in a multi-class classification setting with 15 class targets. The results demonstrate the effectiveness of various machine learning models and techniques. The ensemble model "VGG19" stands out as the top performer across all evaluation metrics, boasting the highest ac- curacy (99.26%), precision (97.93%), recall (99.00%), and F1-Measure (95.53%). This success is attributed to its feature extraction capabilities, transfer learning, robustness, and ability to generalize across different types of network attacks con- tribute to its success in intrusion detection tasks. Notably, the "CNN+LSTM" model also demonstrates strong performance, with high accuracy (97.86%) and Re- call (98.53%), making it a competitive alternative. This success is attributed to its ability to capture temporal and spatial features, particularly beneficial for intrusion detection. In contrast, the "RF" model lags behind the others, achieving the lowest performance scores across all metrics. The results highlight the importance of pre- processing techniques, including balanced data, PCA for dimensionality reduction, and hyperparameter optimization (HPO) using Bayesian optimization. The use of data preprocessing collectively leads to improved intrusion detection classification performance.

Table 1. Accuracy and Precision Results for CIC-IDS2017 Dataset

| Classlabel | Accuracy | | | | | Precision | | | | |
|---|---|---|---|---|---|---|---|---|---|---|
| | CNN | LSTM | CNN+LSTM | RF | VGG19 | CNN | LSTM | CNN+LSTM | RF | VGG19 |
| Normal | 95.65% | 95.40% | 98.09% | 99.00% | 99.88% | 98.00% | 100% | 93.00% | 99.00% | 98.00% |
| DoSHulk | 95.55% | 94.45% | 98.95% | 96.00% | 99.84% | 91.00% | 91.00% | 96.00% | 99.00% | 97.00% |
| DDoS | 96.55% | 96.89% | 99.60% | 94.00% | 98.83% | 98.00% | 100% | 100% | 95.00% | 98.00% |
| PortScan | 97.39% | 95.45% | 96.31% | 97.00% | 98.86% | 91.00% | 99.00% | 97.00% | 94.00% | 99.00% |
| DoSGoldenEye | 86.07% | 97.74% | 97.94% | 95.00% | 96.18% | 90.00% | 100% | 100% | 100% | 99.00% |
| FTP-Patator | 90.40% | 98.03% | 97.62% | 93.00% | 96.49% | 89.00% | 99.00% | 92.00% | 100% | 95.00% |
| DoSslowloris | 91.19% | 99.17% | 99.93% | 67.00% | 99.99% | 87.00% | 99.00% | 93.00% | 98.00% | 98.00% |
| DoSSlowhttptest | 94.68% | 99.42% | 99.88% | 89.00% | 100% | 100% | 99.00% | 97.00% | 99.00% | 99.00% |
| SSH-Patator | 93.68% | 99.48% | 99.94% | 49.00% | 99.9% | 100% | 90.00% | 100% | 99.00% | 97.00% |
| Bot | 92.04% | 91.51% | 93.89% | 84.00% | 99.19% | 93.00% | 85.00% | 100% | 52.00% | 100% |
| WebAttack–BruteForce | 95.19% | 95.30% | 96.48% | 90.00% | 100% | 99.00% | 100% | 100% | 89.00% | 100% |
| WebAttack–XSS | 94.66% | 94.33% | 96.19% | 70.00% | 94.57% | 98.00% | 99.00% | 100% | 95.00% | 95.00% |
| Infiltration | 94.11% | 96.32% | 96.59% | 87.00% | 99.56% | 94.00% | 78.00% | 95.00% | 71.00% | 96.00% |
| WebAttack–SqlInjection | 95.60% | 96.56% | 97.02% | 90.00% | 96.99% | 88.00% | 91.00% | 97.00% | 89.00% | 98.00% |
| Heartbleed | 87.91% | 97.70% | 96.15% | 83.00% | 98.53% | 90.00% | 100% | 91.00% | 95.00% | 98.00% |
| Average | 94.02% | 96.38% | 97.86% | 84.73% | 99.26% | 93.20% | 93.47% | 92.87% | 91.47% | 97.93% |

Table 2. Recall and F1-Measure Results for CIC-IDS2017 Dataset

| Class label | Recall | | | | | F1-Measure | | | | |
|---|---|---|---|---|---|---|---|---|---|---|
| | CNN | LSTM | CNN+LSTM | RF | VGG19 | CNN | LSTM | CNN+LSTM | RF | VGG19 |
| Normal | 99% | 100% | 98.00% | 98.00% | 99.00% | 99.00% | 100% | 99.00% | 99.00% | 99.00% |
| DoSHulk | 84% | 84% | 98.00% | 99.00% | 99.00% | 95.00% | 98.00% | 97.00% | 98.00% | 96.00% |
| DDoS | 100% | 100% | 100% | 100% | 100% | 100% | 96.00% | 97.00% | 98.00% | 98.00% |
| PortScan | 90% | 100% | 100% | 99.00% | 100% | 92.00% | 99.00% | 100% | 69.00% | 97.00% |
| DoSGoldenEye | 99% | 100% | 100% | 100% | 100% | 94.00% | 100% | 96.00% | 100% | 100% |
| FTP-Patator | 88% | 99% | 99.00% | 100% | 99.00% | 93.00% | 99.00% | 96.00% | 100% | 100% |
| DoSslowloris | 91% | 99% | 100% | 98.00% | 100% | 94.00% | 99.00% | 94.00% | 99.00% | 99.00% |
| DoSSlowhttptest | 99% | 99% | 100% | 100% | 100% | 96.00% | 97.00% | 97.00% | 89.00% | 91.00% |
| SSH-Patator | 91.17% | 100% | 100% | 37.00% | 98.00% | 94.00% | 98.00% | 100% | 91.00% | 100% |
| Bot | 100% | 82% | 97.00% | 95.00% | 99.00% | 96.00% | 94.00% | 98.00% | 97.00% | 97.00% |
| WebAttack–BruteForce | 100% | 100% | 98.00% | 98.00% | 93.00% | 100% | 100% | 100% | 98.00% | 99.00% |
| WebAttack–XSS | 98% | 99% | 100% | 89.00% | 98.00% | 96.00% | 98.00% | 97.00% | 97.00% | 89.00% |
| Infiltration | 96.00% | 86% | 93.00% | 85.00% | 89.00% | 93.00% | 96.00% | 96.00% | 88.00% | 95.00% |
| WebAttack–SqlInjection | 85.00% | 88.00% | 95.00% | 91.00% | 93.00% | 96.00% | 97.00% | 99.00% | 93.00% | 94.00% |
| Heartbleed | 95% | 82.00% | 98.00% | 95.00% | 97.00% | 94.00% | 95.00% | 98.00% | 96.00% | 100% |
| Average | 92.15% | 93.87% | 98.53% | 90.07% | 99% | 94.47% | 96.20% | 97.40% | 93.67% | 95.53% |

## 4.3. Results on CIC-IDS2018 Dataset

In the multi-class classification results for intrusion detection using the CSE-CIC- IDS2018 dataset with 15 class targets and balanced data, various performance metrics were evaluated. The results in Table 3 and 4 indicate that the VGG19 model consistently outperforms the other models, achieving the highest Accuracy of 99.22% and F1-Measure of 99.33% on average. This is followed by the CNN+LSTM model with an average Accuracy of 95.82% % , Precision of 98.13% and F1-Measure of 97.00%. The CNN, LSTM, and RF models also performed well, with average ac- curacies of 96.56%, 96.83%, and 89.94%, respectively. However, these models showed slightly lower F1-Measures, indicating that VGG19 and CNN+LSTM models strike a better balance between precision and recall. The hyperparameter optimization of the RF model using Bayesian optimization with tree-based Parzen estimator (BO- TPE) resulted in a competitive performance.

Table 3. Accuracy and Precision Results for CSE-CIC-IDS2018 dataset

| Classlabel | Accuracy | | | | | Precision | | | | |
|---|---|---|---|---|---|---|---|---|---|---|
| | CNN | LSTM | CNN+LSTM | RF | VGG19 | CNN | LSTM | CNN+LSTM | RG | VGG19 |
| Normal | 98.34% | 95.42% | 95.04% | 94.89% | 99.88% | 97.00% | 99.00% | 99.00% | 99.00% | 99.95% |
| DDOSattack-HOIC | 98.18% | 96.06% | 96.24% | 92.85% | 97.84% | 100% | 97.00% | 100% | 99.00% | 98.00% |
| DDoSattacks-LOIC-HTTP | 97.27% | 96.62% | 96.81% | 88.77% | 97.85% | 95.00% | 100% | 100% | 96.00% | 98.00% |
| DoSattacks-Hulk | 98.26% | 97.16% | 95.67% | 90.90% | 99.88% | 89.00% | 100% | 95.00% | 55.00% | 100% |
| Bot | 98.35% | 97.11% | 94.19% | 88.28% | 98.83% | 94.00% | 98.00% | 100% | 96.00% | 100% |
| FTP-BruteForce | 96.31% | 97.67% | 94.51% | 94.47% | 98.46% | 99.00% | 93.00% | 97.00% | 90.00% | 90.00% |
| SSH-BruteForce | 98.39% | 97.63% | 94.33% | 84.12% | 99.19% | 91.00% | 91.00% | 100% | 95.00% | 93.00% |
| Infiltration | 97.42% | 88.26% | 94.01% | 94.12% | 98.99% | 99.00% | 100% | 100% | 89.00% | 99.00% |
| DoSattacks-SlowHTTPTest | 90.63% | 97.93% | 95.31% | 94.29% | 99.68% | 98.00% | 96.00% | 99.00% | 91.00% | 100% |
| DoSattacks-GoldenEye | 86.64% | 98.48% | 96.70% | 85.56% | 99.92% | 95.00% | 92.00% | 95.00% | 85.00% | 100% |
| DoSattacks-Slowloris | 90.78% | 94.04% | 97.23% | 89.70% | 99.95% | 89.00% | 98.00% | 89.00% | 94.00% | 100% |
| DDOSattack-LOIC-UDP | 91.39% | 94.53% | 96.16% | 92.87% | 94.57% | 93.00% | 95.00% | 91.00% | 87.00% | 99.00% |
| BruteForce-Web | 94.24% | 93.73% | 97.38% | 79.03% | 98.52% | 91.00% | 96.00% | 95.00% | 88.00% | 100% |
| BruteForce-XSS | 94.32% | 94.21% | 96.16% | 84.12% | 92.41% | 97.00% | 92.00% | 92.00% | 93.00% | 95.00% |
| SQLInjection | 90.09% | 96.31% | 96.46% | 94.87% | 98.35% | 95.00% | 94.00% | 100% | 90.00% | 97.00% |
| Average | 96.56% | 96.83% | 95.82% | 89.94% | 99.22% | 91.40% | 93.27% | 98.13% | 90.60% | 99.93% |

Table 4. Recall and F1-Measure Results for CSE-CIC-IDS2018 dataset

| Classlabel | Recall | | | | | F1-Measure | | | | |
|---|---|---|---|---|---|---|---|---|---|---|
| | CNN | LSTM | CNN+LSTM | RF | VGG19 | CNN | LSTM | CNN+LSTM | RF | VGG19 |
| Normal | 99.00% | 100% | 100% | 98.00% | 99.00% | 99.00% | 99.00% | 99.00% | 98.00% | 99.00% |
| DDOSattack-HOIC | 100% | 98.00% | 100% | 99.00% | 100% | 100% | 97.00% | 100% | 99.00% | 100% |
| DDoSattacks-LOIC-HTTP | 100% | 97.00% | 98.00% | 100% | 100% | 90.00% | 100% | 93.00% | 98.00% | 93.00% |
| DoSattacks-Hulk | 93.00% | 96.00% | 100% | 100% | 100% | 91.00% | 100% | 100% | 100% | 100% |
| Bot | 100% | 100% | 95.00% | 100% | 100% | 100% | 99.00% | 100% | 100% | 94.00% |
| FTP-BruteForce | 99.00% | 90.00% | 92.00% | 99.00% | 100% | 96.00% | 100% | 94.00% | 99.00% | 96.00% |
| SSH-BruteForce | 100% | 92.00% | 96.00% | 99.00% | 99.00% | 94.00% | 91.00% | 96.00% | 100% | 94.00% |
| Infiltration | 100% | 93.00% | 98.00% | 100% | 100% | 98.00% | 77.00% | 100% | 98.00% | 100% |
| DoSattacks-SlowHTTPTest | 100% | 95.00% | 100% | 99.00% | 97.00% | 97.00% | 95.00% | 98.00% | 100% | 98.00% |
| DoSattacks-GoldenEye | 94.00% | 100% | 87.00% | 100% | 100% | 100% | 89.00% | 100% | 70.00% | 100% |
| DoSattacks-Slowloris | 95.00% | 96.00% | 100% | 43.00% | 93.00% | 77.00% | 79.00% | 88.00% | 100% | 98.00% |
| DDOSattack-LOIC-UDP | 89.00% | 94.00% | 98.00% | 18.00% | 96.00% | 100% | 97.00% | 99.00% | 79.00% | 100% |
| BruteForce-Web | 100% | 92.00% | 94.00% | 89.00% | 89.00% | 92.00% | 98.00% | 90.00% | 93.00% | 100% |
| BruteForce-XSS | 91.00% | 91.000% | 99.00% | 53.00% | 94.00% | 95.00% | 91.00% | 93.00% | 94.00% | 99.00% |
| SQLInjection | 100% | 97.00% | 97.00% | 91.00% | 100% | 100% | 95.00% | 96.00% | 90.00% | 100% |
| Average | 99.00% | 94.53% | 96.93% | 74.19% | 98.47% | 97.33% | 91.40% | 97.00% | 93.20% | 99.33% |

## 4.4. Comparison of Hyper parameter optimization (HOP) Results

### 4.4.1. Comparison of HOP for CIC-IDS2017 Dataset

In Table 5, we analyse the performance of five machine learning algorithms for intrusion detection using the CIC-IDS2017 NetFlow dataset. We compare results in terms of training time, accuracy, and F1-measure, both without and with Hyperparameter Optimization (HOP) using Bayesian optimization with tree-based Parzen estimator (BO-TPE). Without HOP, the CNN classifier achieved 93.72% accuracy in 156 seconds, while the LSTM classifier reached 96.01% accuracy in 572 seconds. The CNN+LSTM model excelled with 96.99% accuracy and 235 seconds of training time. The RF classifier had a longer training time of 1,606 seconds, achieving 83.36% accuracy. The VGG16 model achieved the highest accuracy of 98.55% in 593 seconds. With HOP, the CNN classifier achieved 94.02% accuracy in 130 seconds, and the LSTM classifier achieved 96.38% accuracy in 480 seconds. The CNN+LSTM model displayed the highest accuracy at 97.86% with a significantly reduced training time of 85 seconds. The RF classifier required 1,162 seconds and reached 84.73% accuracy. The VGG16 model achieved the highest accuracy of 99.26% in 534 seconds.

Table 5. Classifier Performance with and without HOP for CIC-IDS2017 dataset.

| Data | Classifier | TrainingTime(s) | Accuracy | F1-Measure |
|---|---|---|---|---|
| W/oHOP | CNN | 156s | 93.72% | 93.33% |
|  | LSTM | 572s | 96.01% | 95.77% |
|  | CNN+LSTM | 235s | 96.99% | 97.09% |
|  | RF | 1,606s | 83.36% | 92.13% |
|  | VGG16 | 593s | 98.55% | 95.11% |
| W/HOP | CNN | 130s | 94.02% | 94.47% |
|  | LSTM | 480s | 96.38% | 96.20% |
|  | CNN+LSTM | 85s | 97.86% | 97.40% |
|  | RF | 1,162s | 84.73% | 93.67% |
|  | VGG16 | 534s | 99.26% | 95.53% |

### 4.4.2. Comparison of HOP for CSE-CIC-IDS2018 Dataset

In Table 6, we present a comparison of results regarding training time, accuracy, and F1-measure, both with and without Hyperparameter Optimization (HOP) using Bayesian optimization with a tree-based Parzen estimator (BO-TPE). Without HOP, the CNN classifier achieved an accuracy of 95.13% with a training time of 2076 seconds, while the LSTM classifier reached 96.04% accuracy but required a longer training duration of 2989 seconds. The CNN+LSTM model obtained 94.22% accuracy with a training time of 239 seconds. The RF classifier had a training time of 553 seconds and achieved 88.00% accuracy. The VGG16 model exhibited the highest accuracy at 98.57% with a training time of 630 seconds. With HOP applied, the CNN classifier achieved 96.56% accuracy with a reduced training time of 1149 seconds. The LSTM classifier reached 96.83% accuracy with a training time of 2924 seconds. The CNN+LSTM model achieved 95.82% accuracy with a significantly reduced training time of 182 seconds. The RF classifier had a training time of 716 seconds and an accuracy of 89.94%. The VGG16 model attained the highest accuracy of 99.22% with a training time of 524 seconds. Overall, the results indicate that the CNN+LSTM model consistently performed well in terms of accuracy and training time, outperforming other models in most cases. While the VGG16 model achieved high accuracy, it required a longer training time.

Table 6. Classifier Performance with and without HOP using CSE-CIC-IDS2018

| Data | Classifier | TrainingTime(s) | Accuracy | F1-Measure |
|---|---|---|---|---|
| W/oHOP | CNN | 2076s | 95.13% | 96.15% |
|  | LSTM | 2989s | 96.04% | 89.20% |
|  | CNN+LSTM | 239s | 94.22% | 96.89% |
|  | RF | 553s | 88.00% | 91.35% |
|  | VGG16 | 630s | 98.57% | 98.79% |
| W/HOP | CNN | 1149s | 96.56% | 97.33% |
|  | LSTM | 2924s | 96.83% | 91.40% |
|  | CNN+LSTM | 182s | 95.82% | 97.00% |
|  | RF | 716s | 89.94% | 93.20% |
|  | VGG16 | 524s | 99.22% | 99.33% |

## 4.5. Comparison with Existing Methods

In comparing existing approaches for intrusion detection, in Table 7 some models have been evaluated on different datasets. The BCNN model, trained on the UNSW-NB15 data set in 2021, achieved an accuracy of 90.25% and an F1-measure of 90.45%. The Tree-CNN model, trained on the CIC-IDS2017 data set in 2021, demonstrated higher performance with an accuracy of 98.00% and an F1-measure of 98.00%. The OHDNN model, trained on the UNSW-NB15 dataset in 2023, achieved an accuracy of 98.30% and an F1-measure of 97.10%. Finally, the proposed model, trained on the CSE-CIC IDS2018 dataset in 2023, utilized the VGG19 architecture and achieved even higher accuracy and F1-Measure, with results of 99.22% and 99.33% respectively. The proposed model outperforms the other approaches in terms of accuracy, show casing the effectiveness of the VGG19 model for intrusion detection.

Table 7. Comparison of Models' Performance

| Model | Dataset | Year | Accuracy | F1-Measure |
|---|---|---|---|---|
| BCNN[22] | UNSW-NB15 | 2021 | 90.25% | 90.45% |
| Tree-CNN[23] | CIC-IDS2017 | 2021 | 98.00% | 98.00% |
| OHDNN+ECRF[24] | UNSW-NB15 | 2023 | 98.30% | 97.10% |
| PCA-DNN [25] | CSE-CICIDS2018 | 2022 | 96.79% | - |
| Proposed | CIC-IDS2017 | 2023 | VGG19=99.26% | VGG19=95.53% |
| | CSE-CIC IDS2018 | 2023 | VGG19=99.22% | VGG19=99.33% |

## 5. CONCLUSIONS

In this paper, we present an advanced intrusion detection solution that combines deep learning and processing techniques, addressing challenges related to imbalanced data, high-dimensional feature spaces, and optimized hyperparameters to reduce false positives. Our performance analysis on the CIC-IDS2017 and CSE- CIC-IDS2018 datasets highlight the effectiveness of our approach. The ensemble model "VGG19" consistently outperforms other models, demonstrating high ac- curacy, precision, recall, and F1-Measure. This success is attributed to VGG19's feature extraction, transfer learning, robustness, and generalization capabilities. The "CNN+LSTM" model also performs well by capturing both temporal and spatial features. However, the "RF" model lags behind, emphasizing the importance of model selection in intrusion detection. In future work, we plan to enhance our intrusion detection system by exploring dynamic ensemble strategies that adapt to evolving attack patterns. Additionally, we aim to investigate the integration of anomaly detection techniques to further boost the system's ability to identify novel threats.

## ACKNOWLEDGEMENTS

The authors would like to thank everyone, just everyone!